\documentclass{JHEP3}
\usepackage{epsfig}

\title{AGT on the S-duality Wall}

\author{Kazuo Hosomichi\\
	Yukawa Institute for Theoretical Physics\\
	E-mail: \email{hosomiti@yukawa.kyoto-u.ac.jp}}

\author{Sungjay Lee\\
	Korea Institute for Advanced Study;\\
    DAMTP, Centre for Mathematical Sciences, University of Cambridge,\\
     Wilberforce Road, Cambridge, CB3 0WA, UK \\
	E-mail: \email{sjlee@kias.re.kr, \\\,\,\,\,\, \,\,\,\,  s.lee@damtp.cam.ac.uk}}

\author{Jaemo Park\\
	Department of Physics \&
        Postech Center for Theoretical Physics, POSTECH\\
	E-mail: \email{jaemo@postech.ac.kr}}

\abstract{
Three-dimensional gauge theory $T[G]$ arises on a domain wall
between four-dimensional ${\cal N}=4$ SYM
theories with the gauge groups $G$ and its S-dual $G^{\rm L}$.
We argue that the ${\cal N}=2^\ast$ mass deformation of the bulk theory
induces a mass-deformation of the theory $T[G]$ on the wall.
The partition functions of the theory $T[SU(2)]$ and its
mass-deformation on the three-sphere are shown to coincide
with the transformation coefficient of Liouville one-point conformal
block on torus under the S-duality.
}

\preprint{KIAS-P10025\\YITP-10-76}


\begin{document}

\section{Introduction}

Gauge theories with eight supersymmetry in various dimensions
are interesting playgrounds in which one can test by rigorous
computations our intuitive understandings of physics at strong
gauge coupling. One remarkable achievement in recent years was the exact
computation of partition function of 4D ${\cal N}=2$
gauge theories on Omega-background \cite{Nek} which made use of
the localization techniques.
More recently it was found that the same idea of localization applies
to the computation of partition function of gauge theories on
$S^4$ \cite{Pe} or $S^3$ \cite{KWY1, KWY2}.
These exact results have also led to a discovery of a surprising
relationship between 4D gauge theories and 2D CFTs, which we call
AGT relation.
A correspondence between $SU(2)$ quiver gauge theories and Liouville
theory has been found by Alday, Gaiotto and Tachikawa (AGT) \cite{AGT},
and it has been generalized by Wyllard \cite{Wy}
to higher rank gauge theories and Toda theories.

After the discovery of this remarkable relation, various generalization
have been considered to include external charged objects or topological
defects.
As an example, a correspondence was proposed in \cite{AGGTV,DGOT}
between Wilson or 't Hooft loops in gauge theory and Verlinde's loops
or the loops of topological defect in 2D theories, and it was
studied further in \cite{Pas,GF}. It was also argued in \cite{AGGTV}
that the effect of some external vortex strings or
surface operators in gauge theories can be identified with that of
degenerate field operators in 2D theories (see also \cite{MT}),
and the correspondence has been verified explicitly in \cite{AFKMY}.
There are other surface operators which, according to \cite{AT,KPPW},
changes the 2D theories from Liouville or Toda theories to those
with affine Lie algebra symmetry.

More recently, several suggestions have been made in \cite{DGG}
on the possible correspondences between external domain walls in
4D gauge theories and certain topological defect operators in 2D theories.
Of particular interest are the configurations with a Janus domain wall
where the gauge couplings on the two sides of the wall are related
by a generalized S-duality group action.
By taking the S-duality on one side of the wall, we obtain the
{\it S-duality wall} separating two gauge theories which are
S-dual to each other.
By studying such domain walls one is led to expect
that the partition function of the 3D field theory on the wall
should coincide with the S-duality transformation
coefficients of conformal blocks, which we call
``S-duality kernel'' for short.

In this paper we focus on a 3D gauge theory called $T[SU(N)]$
which arises on the S-duality wall in ${\cal N}=4$ $SU(N)$ SYM theory.
We start with the simplest theory $T[SU(2)]$ and identify its mass-deformation
corresponding to the bulk ${\cal N}=2^\ast$ deformation.
Then we show the agreement up to some normalization factors between
the partition function of the mass-deformed theory on $S^3$ and
the S-duality kernel of Liouville conformal blocks on one-punctured
torus under the S-duality action
which maps the complex structure of the torus from $\tau$ to $-1/\tau$.
By a simple generalization of our argument, one can find the mass-deformation
of $T[SU(N)]$ theories with $N\ge3$ corresponding to the bulk
${\cal N}=2^\ast$ deformation.
We conjecture that the partition function of the mass-deformed $T[SU(N)]$
theory on $S^3$ corresponds to the S-duality kernel of Toda theory
on one-punctured torus, with a certain degenerate operator
inserted at the puncture.

\section{Reviews}

Here we begin by reviewing recent works on $1/2$ BPS boundaries and
domain walls of ${\cal N}=4$ SYM theory and the 3D superconformal
theory $T[G]$. Then we move on to the AGT relation and explain
how it is generalized in the presence of Janus domain walls.
These are necessary to formulate the conjecture on the S-duality wall
which will be tested later on.

\subsection{3D theory $T{[G]}$ on boundaries and domain walls in ${\cal N}=4$ SYM}

Boundary conditions on 4D ${\cal N}=4$ SYM have been
classified in a recent paper \cite{GW1}, and their property under S-duality
transformation has been extensively discussed in \cite{GW3}.
It was shown in \cite{GW3} that the S-dual of Dirichlet boundary
condition on the ${\cal N}=4$ SYM with gauge group $G$
(more precisely, Dirichlet condition on
${\cal N}=2$ vectormultiplet fields and Neumann condition on
hypermultiplet fields) is described by a boundary field theory
$T[G]$ coupled to the bulk gauge field.

$T[G]$ is a 3D ${\cal N}=4$ gauge theory with global
symmetry $G\times G^{\rm L}$ (Langlands dual of $G$) at IR
superconformal fixed point. It has two distinct vacuum moduli spaces,
namely Higgs and Coulomb branches, which admit the action of $G$ and
$G^{\rm L}$ respectively. Under the ${\cal N}=4$ mirror transformation,
$T[G]$ is mapped to $T[G^{\rm L}]$.

For $G=SU(N)$, there is a simple type IIB brane construction for $T[SU(N)]$.
We realize the $SU(N)$ SYM on $N$ coincident D3-branes stretching
along the directions $0126$ with $x_6\ge0$. In order to implement
ordinary Dirichlet boundary condition without any Nahm pole structure
at $x_6=0$, we need $N$ D5-branes extending along the directions
$012345$ at $x_6=0$ so that each D3-brane ends on different D5-branes.

We take the S-dual of the SYM theory which maps the D5-branes into
NS5-branes. See Figure \ref{tg}(a).
The resulting brane configuration gives rise to a 3D ${\cal N}=4$
gauge theory on the boundary which is characterized by the following
quiver diagram.
\[
 \epsfig{file=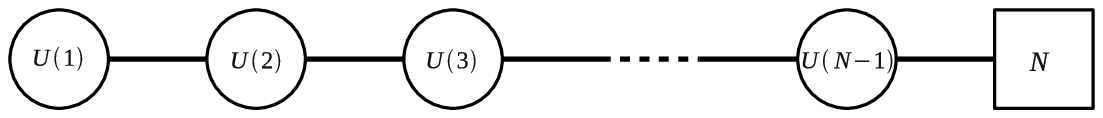, width=100mm}
\]
The 3D gauge group $\otimes_{n=1}^{N-1}U(n)$ and global symmetry $SU(N)$
correspond to the round and square nodes respectively, and there is a
hypermultiplet corresponding to each link.
This theory couples to the bulk fields through the gauging of $SU(N)$
global symmetry, and defines the S-dual of Dirichlet boundary condition
of ${\cal N}=4$ SYM.

In this S-dual picture, the four-dimensional fields on D3-branes can be
frozen by introducing another $N$ D5-branes stretching along the directions
$012789$ at positive $x_6$ and terminating the D3-branes on them.
See Figure \ref{tg}(b). The resulting brane configuration
defines the 3D theory $T[SU(N)]$.
It is manifestly invariant under S-duality transformation if the spatial
directions $345$ and $789$ are exchanged at the same time.
It follows that the Coulomb and Higgs branch moduli spaces of $T[SU(N)]$
are isomorphic and both admit the action of $SU(N)$.

\FIGURE[bht]{
\epsfig{file=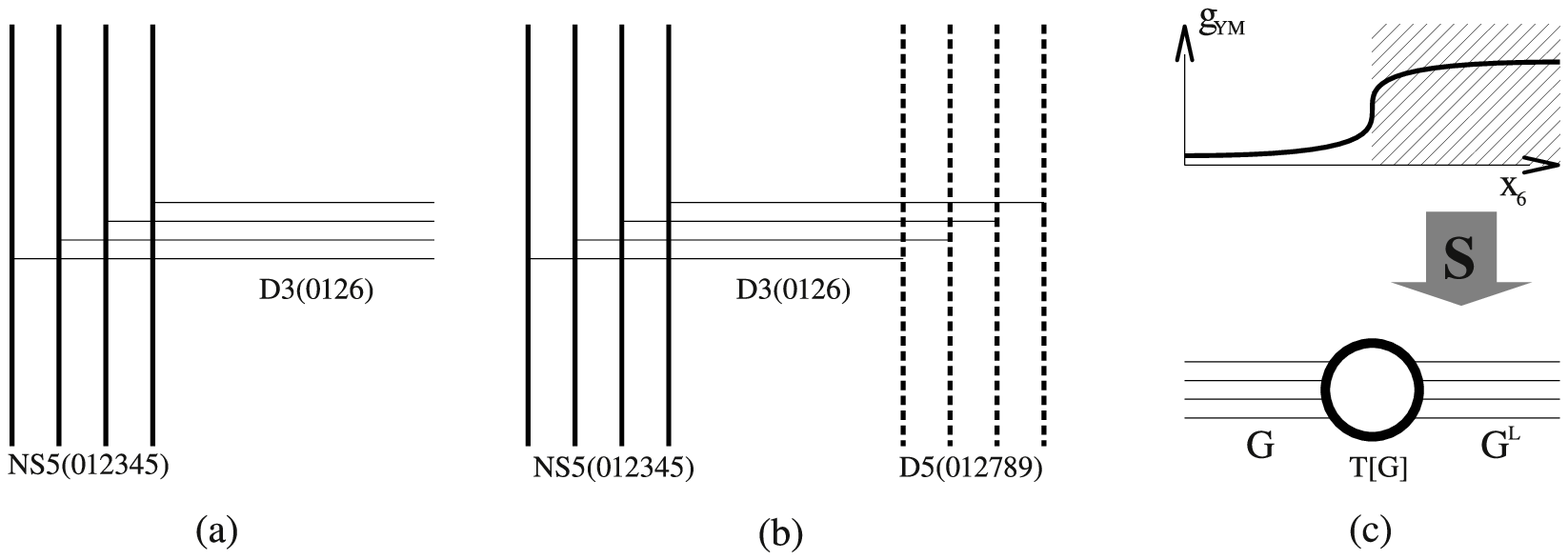, width=140mm}
\caption{
(a) Type IIB brane construction of the 3D gauge theory
$T[SU(N)]$ coupled to 4D ${\cal N}=4$ SYM theory.
(b) Type IIB brane construction for $T[SU(N)]$.
(c) Starting from Janus domain wall of $G$ SYM theory and S-dualizing
only the right of the wall, one obtains $T[G]$ connecting $G$ and
$G^{\rm L}$ SYM theories.}
\label{tg}
}

It was also argued in \cite{GW3} that the theory $T[G]$ arise in
relation to Janus domain walls of ${\cal N}=4$ SYM connecting the
same theory at different values of coupling. In the presence of
such domain walls, the holomorphic gauge coupling $\tau$ varies as
a function of one of the coordinates, say $x_6$.
Suppose the gauge coupling is $\tau$ on the left of the wall ($x_6<0$)
and $-1/\tau$ on the right ($x_6>0$). If the theory on the left is
weakly coupled, it is strongly coupled on the right. By taking the
S-dual only on the right half space $x_6>0$, one obtains a wall
connecting the $G$ SYM on the left and $G^{\rm L}$ SYM on the right,
with the same coupling $\tau$ being small on both sides.
This is called the S-duality wall. See Figure \ref{tg}(c).

Note that, if we introduce the Dirichlet boundary on $G^{\rm L}$ SYM at
some positive $x_6$ and take the S-duality of the region $x_6>0$ again,
the system flows down to the $G$ SYM on left half-space with the
S-dual of Dirichlet boundary for $G^{\rm L}$ SYM.
It follows that the 3D theory on the S-duality wall is nothing but
$T[G]$ coupled to the bulk fields through the gauging of its
$G\times G^{\rm L}$ global symmetry.

As the simplest example, $T[SU(2)]$ is a ${\cal N}=4$ SQED with two
electron hypermultiplets which is long known to be self-dual \cite{IS,KS}.
In this letter we will mainly focus on this theory and its
mass-deformation to ${\cal N}=2$.

\subsection{AGT relation}

In a recent work \cite{AGT} a remarkable correspondence was found
between partition functions of certain ${\cal N}=2$ superconformal
gauge theories on $S^4$ \cite{Pe} and correlation functions of
Liouville theory with central charge $c=25$, Liouville coupling $b=1$.
Both can be written in the following form,
\begin{equation}
 Z ~=~ \int d\nu(\alpha) \
 \overline{\cal F}^{(\sigma)}_{\alpha,E}(\tau)
 {\cal F}^{(\sigma)}_{\alpha,E}(\tau).
\end{equation}
Let us explain the ingredients of the formula for each side
of the correspondence.

In Liouville theory, $Z$ is the correlation function of primary
operators on a Riemann surface $\Sigma$. We denote its complex
structure by $\tau$, and choose an arbitrary pants decomposition $\sigma$.
We also think of the corresponding Moore-Seiberg graph drawn on
$\Sigma$, which consists of trivalent vertices and legs.
The external legs originate from the punctures and are assigned the
Liouville momenta $E=\{E_i\}$, and similarly $\alpha=\{\alpha_a\}$
is the collection of Liouville momenta assigned to the internal edges.
${\cal F}$ and $\overline{\cal F}$ are the Virasoro conformal blocks for
the holomorphic and anti-holomorphic sectors.
To construct correlation functions, one takes the product of conformal
blocks and integrate over $\alpha$ with a measure $d\nu(\alpha)$
which is uniquely determined from single-valuedness and other
consistency conditions.
Among those conditions is the requirement that, although the conformal
blocks are defined with respect to a specific pants decomposition $\sigma$,
the correlation functions do not depend on the choice of $\sigma$.
This is a strong requirement since the Liouville conformal blocks are
subject to nontrivial integral transformations of the form
\begin{equation}\label{integraltransform}
 {\cal F}^{(\sigma)}_{\alpha,E}(\tau)~=~
 \int d\nu(\alpha') \
 g^{(\sigma,\sigma')}_{(\alpha,\alpha',E)}
 {\cal F}^{(\sigma')}_{\alpha',E}(\tau),
\end{equation}
under changes of pants decomposition from $\sigma$ to $\sigma'$.
The coefficient $g^{(\sigma,\sigma')}_{(\alpha,\alpha',E)}$ is
called the S-duality kernel in what follows.

In the gauge theory side, we take a generalized superconformal
quiver theory \cite{Ga} with an $SU(2)$ gauge
group at each node.
Some of such theories have type IIA brane construction, and at
low energy they are described by two M5-branes wrapping
a Riemann surface $\Sigma$ with punctures.
In \cite{Ga} it was shown that there is a family of
ganeralized quiver theories for each punctured Riemann surface $\Sigma$.
Each member of the family has a definite generalized quiver structure
and corresponds to a specific pants decomposition of $\Sigma$,
and different members are related to one another by S-dualities.
The gauge couplings $\tau$ determine
the complex structure of $\Sigma$,
the parameter $E$ determines the masses and $\alpha$ corresponds to
the expectation value of scalars in vectormultiplet of the theory
on $\mathbb{R}^4$.
${\cal F},\overline{\cal F}$ are each identified with the Nekrasov partition
function of the theory on $\mathbb{R}^4$ with the Omega-deformation
parameter $\epsilon_1=\epsilon_2=1$ in unit of $\hbar$.
The integration measure $d\nu(\alpha)$ now represents the
perturbative contribution to the partition function.
By taking the bilinear product of ${\cal F}$ and integrating over
the Coulomb branch moduli space one obtains the partition function
of the gauge theory on $S^4$. It is noteworthy here that
(\ref{integraltransform}) can be understood as the generalized
S-duality action on
the Nekrasov partition function of ${\cal N}=2$ gauge theories.

\subsection{Partition function with domain walls: a conjecture}

Among various generalizations of the AGT relation that have been
proposed so far, we are interested in the inclusion of Janus domain walls.

Based on the origin of $\cal F, \overline F$ in the computation of
\cite{Pe}, it has been proposed in \cite{DGG} that the gauge theory
partition function in the presence of Janus domain wall wrapping on
$S^3$ can be expressed by simply choosing different complex structure
moduli for ${\cal F}$ and $\overline{\cal F}$,
\begin{eqnarray}
  Z ~=~ \int d\nu(\alpha)\ \overline{\cal F}^{(\sigma)}_{\alpha,E}(\tau)
  {\cal F}^{(\sigma)}_{\alpha,E}(\tau').
\end{eqnarray}
As a special case, if the two moduli are not the same but are related
by an element $g$ of the S-duality group as $\tau'=g\cdot\tau$,
then one has
\begin{equation}
 {\cal F}^{(\sigma)}_{\alpha,E}(g\cdot\tau)~=~
 {\cal F}^{(g\cdot\sigma)}_{\alpha,E}(\tau)~=~
 \int d\nu(\alpha')\
 g^{(\sigma)}_{(\alpha,\alpha',E)}{\cal F}^{(\sigma)}_{\alpha',E}(\tau),
\end{equation}
with an S-duality kernel $ g^{(\sigma)}_{(\alpha,\alpha',E)}$.

The partition function in the presence of the S-duality wall, obtained
from the Janus domain wall by applying the S-duality
only on one side of the wall, thus becomes
\begin{equation}
 Z ~=~ \int d\nu(\alpha)d\nu(\alpha')\
 \overline{\cal F}^{(\sigma)}_{\alpha,E}(\tau)
 g_{(\alpha,\alpha',E)}^{(\sigma)}
 {\cal F}^{(\sigma)}_{\alpha',E}(\tau).
\end{equation}
Here ${\cal F},\bar{\cal F}$ arise from the two
hemispheres separated by the wall which are labelled by
the Coulomb branch parameters $\alpha,\alpha'$.
One is then naturally led to propose that the S-duality kernel
for $g$ coincides with the partition fucntion of the 3D theory on
the S-duality wall coupled to given ${\cal N}=2$ gauge theory
\begin{eqnarray}\label{conj}
  g^{(\sigma)}_{(\alpha,\alpha',E)} =
  Z_{\rm 3D}[\alpha,\alpha',E]\ .
\end{eqnarray}
For the specific example of ${\cal N}=4$ SYM with gauge group $G$
and $g$ the modular S-duality of the torus, the 3D theory on the
wall is known to be $T[G]$. The bulk Coulomb branch parameters
$\alpha,\alpha'$ on two sides of the wall become the FI and mass
parameters of the 3D theory and break the global symmetries
$G, G^{\rm L}$ to their Cartan subgroups.
It has been conjectured in \cite{DGG} that the partition function
of $T[G]$ should agree with the modular S-matrix element of the Toda
characters (zero-point blocks on the torus).
As we will see below, the actual correspondence is a little more
involved than that.

\section{Tests on  ${\cal N}=4$/${\cal N}=2^\ast$ $SU(2)$ SYM}

Here we test the proposal (\ref{conj}) concretely in the example of
$SU(2)$ ${\cal N}=4$ SYM as well as its mass-deformation called
${\cal N}=2^\ast$ theory. These theories describe the dynamics of
two M5-branes wrapping $T^2$ with a puncture.
The Nekrasov partition function for the theory on $\mathbb{R}^4$ is
a function of the gauge coupling $\tau$, the mass $m$ for adjoint
hypermultiplet and the expectation value $a$ of the vectormultiplet
scalar.
(It also depends on the parameters of Omega-deformation,
$\epsilon_1=\epsilon_2=\hbar$.)
The corresponding Virasoro conformal block is the one-point block on
a torus of modulus $\tau$, and is described by the Moore-Seiberg
graph of Figure \ref{ms}.
The representation of Virasoro algebra labelled by $p_i$ corresponds
to the Liouville vertex operator $\exp(Q+2ip_i)\phi$, and is of
conformal weight $p_i^2+Q^2/4$.
The labels $p_a$ and $p_e$ are associated to the loop and the external
leg of the Moore-Seiberg graph. They are related to $a,m$ in a certain way,
as we will find out below.

\FIGURE[bht]{
\epsfig{file=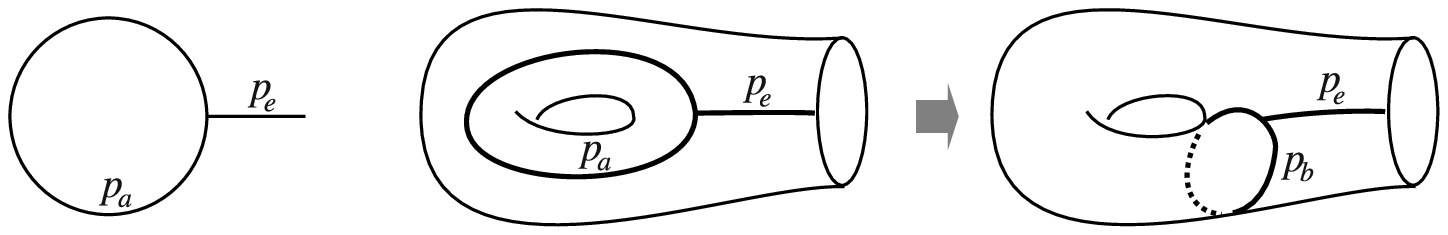, width=140mm}
\caption{(left) Moore-Seiberg graph for one-point block on the torus.
(right) The action of S-duality on Moore-Seiberg graph on one-punctured torus.}
\label{ms}
}

We introduce a Janus domain wall so that the gauge couplings on the
left and right of the wall, $\tau$ and $-1/\tau$, are related by the
S-duality.
The partition function in the presence of this wall is a
product of conformal blocks $\overline{{\cal F}}_{p_a,p_e}(\tau)$,
${\cal F}_{p_b,p_e}(\tau)$ and the S-duality kernel $S_{(p_a,p_b,p_e)}$
integrated over $p_a,p_b$.
The conjecture (\ref{conj}) is that this kernel agrees with the
partition function of a suitable mass deformation of the 3D theory
$T[SU(2)]$.

The relevant S-duality kernel has been obtained in \cite{T3} by making use
of the relationship between Liouville conformal blocks and wave
functions in quantum Teichm\"uller theory \cite{T1,T2,T3,T4}.
\begin{equation}\label{teschner}
 S_{(p_a,p_b,p_e)} = \frac{2^{\frac32}}{s_b(p_e)}
 \int_{\mathbb{R}}dr
 \frac{s_b(p_b+r+\frac12p_e+\frac{iQ}4)}
      {s_b(p_b+r-\frac12p_e-\frac{iQ}4)}
 \frac{s_b(p_b-r+\frac12p_e+\frac{iQ}4)}
      {s_b(p_b-r-\frac12p_e-\frac{iQ}4)}
 e^{4\pi ip_ar}.
\end{equation}
The special function $s_b(x)$ appearing here is characterized by
the normalization condition $s_b(0)=1$, the unitarity condition
$s_b(x)s_b(-x)=1$ and the poles at
$x=\frac{iQ}2+imb+inb^{-1}~(m,n\in\mathbb{Z}_{\ge0})$.
It has an infinite product representation
\begin{equation}
 s_b(x)~=~ \prod_{m,n\in\mathbb{Z}_{\ge0}}
 \frac{mb+nb^{-1}+\frac Q2-ix}{mb+nb^{-1}+\frac Q2+ix}.
\end{equation}
For more detailed explanation on this function, we refer to \cite{KLS,BT}.

When $\frac Q2+ip_e\equiv 2\delta\to 0$, the external momentum vanishes
and the conformal block ${\cal F}_{p_a,p_e}(\tau)$ reduces to the
Virasoro character.
The S-duality kernel also simplifies in this limit to
\begin{equation}
 S_{(p_a,p_b,p_e)}~=~
 \frac{\sqrt2\cos(4\pi p_ap_b)}{\sinh(2\pi b p_b)\sinh(2\pi p_b/b)}.
\label{modS}
\end{equation}
To see how this happens, notice first that the factor $1/s_b(p_e)$ turns
to vanish in this limit. It is cancelled by a divergence arising from
the $r$-integral over the regions $p_b=\pm r$, where two poles pinch
the integration contour.
The integral near $r\simeq p_b$ takes the form
\begin{equation}
 S_{(p_a,p_b,p_e)}~=~ 2^{\frac32}\cdot 4\pi\delta\int dr
 \frac{e^{4\pi ip_ar}}{4\sinh(2\pi b p_b)\sinh(2\pi p_b/b)}
 \frac{1}{4\pi^2}\frac1{(r-p_b)^2+\delta^2}+\cdots.
\end{equation}
The integrand becomes proportional to delta function in the limit
$\delta\to0$. There is a similar contribution from the region $r\simeq -p_b$,
and by adding the two contributions we obtain (\ref{modS}).

One can determine the measure of integration over $p_a, p_b$
from the requirement that S-duality operation squares to identity.
\begin{equation}
 d\nu(p)~\equiv~ dp\sinh(2\pi pb)\sinh(2\pi p/b).
\end{equation}
The factors of $\sinh$ functions cancel the denominator of
(\ref{modS}), and one is left with the modular S-matrix element
for non-degenerate Virasoro representations.

\subsection{${\cal N}=4$ SYM and $T{[SU(2)]}$}

The 3D theory $T[SU(2)]$ is an ${\cal N}=4$ SQED with two electron
hypermultiplets.
In terms of ${\cal N}=2$ superfields, this theory has one abelian
vector multiplet $V$, a neutral chiral multiplet $\phi$
and four chiral superfields $q_1,q_2,\tilde q^1,\tilde q^2$
with charge $+1,+1,-1,-1$.
The Lagrangian takes the following form
\begin{eqnarray}
  {\cal L}~=~ \int d^4\theta \ \frac{1}{g^2}\Big[ - \Sigma^2 + |\phi|^2 \Big]
  + \Big[ q^{i\dagger}  e^{-2V}q_i + {\tilde q_i}^\dagger e^{+2V} \tilde q^i \Big]
  + \Big[ \int d^2\theta \  \sqrt2 \tilde q^i \phi q_i + {\rm c.c.} \Big]\ .
\end{eqnarray}
The theory has an $SU(2)_f$ flavor symmetry
which rotates $q_i$ as fundamantal and $\tilde q^i$ as anti-fundamental
doublet. Under its $U(1)$ subgroup the four chiral matters carry charges
$+1,-1,-1,+1$.
The expectation value of the ${\cal N}=2$ vector multiplet scalar is
denoted by $\sigma$.
It becomes the only parameter for the saddle points in the localization
computation of partition function on $S^3$.

One can introduce real masses $\mu$ to the matter fields and
lift the Higgs branch by weakly gauging the flavor symmetry
$U(1) \subset SU(2)_f$ by a background vector
superfield $V_{\rm mass}= - i \theta \bar \theta \mu$,
\begin{eqnarray}
  {\cal L}_{\rm mass} ~=~
  \int d^4 \theta \ \Big[ q^{1\dagger} e^{-2 V_{\rm mass}} q_1 +
  q^{2\dagger} e^{+2 V_{\rm mass} } q_2 +
  \tilde q_1^\dagger e^{+ 2 V_{\rm mass}} \tilde q^1 +
  \tilde q_2^\dagger e^{-2 V_{\rm mass}} \tilde q^2 \Big] \ .
\end{eqnarray}
One can also lift the Coulomb branch by introducing the Fayet-Iliopoulos
(FI) parameter $\zeta$, or in other words by weakly gauging the shift
symmetry of dual photon by another background vector superfield
$V_{\rm FI} = - i \theta \bar \theta \zeta$,
\begin{eqnarray}
  {\cal L}_{\rm FI} ~=~ -\frac{4}{\pi} \int d^4\theta \ V_{\rm FI} \Sigma
  ~=~ - \frac{2}{\pi} \zeta D\ .
\end{eqnarray}
Here the normalization $\frac{4}{\pi}$ is chosen for later convenience.
If this theory appears on the S-duality wall of ${\cal N}=4$ SYM theory
with the gauge group $SU(2)$, these background vector multiplets
$(V_{\rm mass},\phi_{\rm mass})$ and $(V_{\rm FI},\phi_{\rm FI})$
can be identified with bulk vector multiplets on the two sides of the wall.
In particular, Coulomb branch parameters on the two sides of the wall
are identified with $\mu$ and $\zeta$ after a suitable
$SU(2)_N\times SU(2)_R$ R-symmetry rotation.

The partition functions of general 3D ${\cal N}=2$ gauge theory on
$S^3$ have been analyzed in \cite{KWY1,KWY2}, and their
result immediately applies to our problem.
The partition function is an integral over the Coulomb branch moduli
$\sigma$, and the integrand consists of one-loop determinants of gauge
and matter multiplets on the saddle point labeled by $\sigma$.
In our problem, the one-loop determinant of vector multiplet is trivial
but the four matter chiral multiplets yield
\begin{equation}
 Z(\sigma+\mu)Z(\sigma-\mu)Z(-\sigma-\mu)Z(-\sigma+\mu),
\end{equation}
where \cite{KWY1}
\begin{equation}
 Z(\sigma)~\equiv~ \prod_{n=1}^\infty
 \left(\frac{n+\frac12+i\sigma}{n-\frac12-i\sigma}\right)^n ~=~
 s_{b=1}(\textstyle\frac i2-\sigma).
\end{equation}
The FI coupling gives rise to another factor $e^{4\pi i\sigma\zeta}$
in the integrand. The partition function of the defect theory $T[SU(2)]$
coupled to ${\cal N}=4$ SYM is thus given by
\begin{eqnarray}
  Z_{\rm 3D}^{{\cal N}=4} = \int d\sigma \
  \frac{s_{b=1}(\mu+\sigma+\frac i2)}{s_{b=1}(\mu+\sigma-\frac i2)}
  \frac{s_{b=1}(\mu-\sigma+\frac i2)}{s_{b=1}(\mu-\sigma-\frac i2)}
   e^{4\pi i \zeta \sigma}\ ,
\end{eqnarray}
up to a normalization constant.

By a straightforward comparison, one finds that the partition function
agrees with the S-duality kernel (\ref{teschner}) under the
identification of the parameters
\begin{equation}
 b=1,\quad
 r=\sigma,\quad
 p_a=\zeta,\quad
 p_b=\mu,\quad
 p_e=0.
\end{equation}
The last relation implies that, against our intuition, the Liouville
conformal block correspoding to ${\cal N}=4$ $SU(2)$ SYM should
be a torus one-point block with a non-vanishing external momentum,
$\frac Q2+ip_e\ne0$.
One would naturally ask if this value has some special meaning.
In \cite{KWY1} it has been shown that the one-loop determinant of
a general charged ${\cal N}=4$ hypermultiplet can be expressed by
the function
\begin{equation}
 Z(\sigma)Z(-\sigma)~=~ \frac{1}{2\cosh\pi\sigma}.
\end{equation}
We expect that for $b\ne 1$ it will generalize to an identity of
the function $s_b$,
\begin{equation}
  {\textstyle s_b(\frac {ib}2-\sigma)s_b(\frac {ib}2+\sigma)}
 ~=~ \frac{1}{2\cosh\pi b\sigma},
\label{sbcosh}
\end{equation}
or similar identity with $b$ replaced by $1/b$.
If this is the case, then the value of the external momentum for
general $b$ should be that of the Liouville interaction operators
\[
 \textstyle\frac Q2+ip_e=b^{-1}~~\mbox{or}~~ b.
\]
This is in accordance with the observation of \cite{OP}.

\subsection{${\cal N}=2^\ast$ SYM and mass-deformed $T{[SU(2)]}$}

The above result implies that the ${\cal N}=4$ SYM corresponds to the
Liouville theory on a one-punctured torus with external momentum $p_e=0$.
Other values of $p_e$ should correspond to the mass-deformation to
${\cal N}=2^\ast$ theory. What 3D theory should arise on the S-duality
wall of the ${\cal N}=2^\ast$ theory?

The mass deformation of the bulk 4D ${\cal N}=4$ SYM
theory will induce a mass deformation of $T[SU(2)]$ on the wall
which preserves the ${\cal N}=2$ supersymmetry as well as
$SU(2)\times SU(2)$ global symmetries.
Due to these symmetry constraints, the deformation of the theory on the
wall is easily identified as the real mass deformation by weakly gauging
a $U(1)$ symmetry under which $q_1, q_2, \tilde q^1, \tilde q^2$
all carry the same charge $+1$ and $\phi$ carries the charge $-2$.
The deformation to the Lagrangian is
\begin{eqnarray}
  {\cal L}_{\rm def} =
  \int d^4 \theta \ \phi^\dagger e^{4V_{\rm def}} \phi
  + \sum_{i=1}^2 \Big[ q^{i\dagger} e^{-2 V_{\rm def}} q_i
  +\tilde q_i^\dagger e^{- 2 V_{\rm def}} \tilde q^i \Big]  \ ,
\end{eqnarray}
where $V_{\rm def}=im\theta \bar \theta/2$.

As was explained in \cite{Tong}, this $U(1)$ is an anti-diagonal
sum of the R-symmetries $U(1)_N$ and $U(1)_R$ which are
subgroups of the $SU(2)_N\times SU(2)_R$ R-symmetry group of the
undeformed ${\cal N}=4$ theory.
Since $SU(2)_N$ and $SU(2)_R$ are interchanged under the mirror
symmetry (or S-duality in the bulk), the mass-parameter
$m$ is mapped to minus itself under the S-duality.

Applying the localization technique again, one can show that
the one-loop determinant of four chiral multiplets is now modified to
\begin{equation}
\textstyle
 Z( \sigma+\mu-\frac m2)Z( \sigma-\mu-\frac m2)
 Z(-\sigma-\mu-\frac m2)Z(-\sigma+\mu-\frac m2).
\end{equation}
If the contribution from the fields in the ${\cal N}=4$ vector
multiplet $(V,\phi)$ remains trivial, the partition function of the
mass-deformed $T[SU(2)]$ is given by
\begin{eqnarray}\label{n=2}
  Z_{\rm 3D}^{{\cal N}=2^\ast}  = \int d\sigma \
  \frac{s_{b=1}(\mu+\sigma+\frac m2+\frac i2)}
       {s_{b=1}(\mu+\sigma-\frac m2-\frac i2)}
  \frac{s_{b=1}(\mu-\sigma+\frac m2+\frac i2)}
       {s_{b=1}(\mu-\sigma-\frac m2-\frac i2)}
  e^{4\pi i \zeta \sigma}\ .
\end{eqnarray}
By carefully relating the external Liouville momentum $p_e$ and
${\cal N}=2^\ast$ mass parameter $m$, one finds that when $b=1$ and
\begin{equation}
 m~=~ p_e,
\end{equation}
the partition function (\ref{n=2}) agrees with the S-duality kernel
(\ref{teschner}) up to a factor $s_b(-m)$ which is independent of
$p_a,p_b$.

In gauge theory, an additional factor of $s_b(-m)$ is naively expected
to arise from a neutral chiral matter of mass $-2m$.
We claim that it is precisely the contribution of the neutral chiral
field $\phi$ in ${\cal N}=4$ vectormultiplet after the mass deformation.
The field $\phi$ actually has non-canonical R-charge
$1$, and the one-loop determinant for such chiral matters is not known yet.
It would be interesting to explicitly work it out, but we leave it as
a future problem.
We give one supporting argument for our claim in the next subsection.

\subsection{Self-mirror property}\label{sec:self-mirror}

An interesting observation made in \cite{KWY2} is that the function
$1/\cosh\pi x$ is invariant under Fourier transform, and it was used
in the proof of dualities in 3D ${\cal N}=4$ gauge theories.
Here we consider how this can be generalized to mass-deformed
${\cal N}=2$ theories.

We begin by recalling that the function $s_b$ is characterized by
the (\ref{sbcosh}) as well as $s_b=s_{1/b}$ and $s_b(x)s_b(-x)=1$.
Let us define
\begin{equation}
F_{m,b}(x)\equiv \frac{s_b(x+\frac m2+\frac{iQ}4)}{s_b(x-\frac m2-\frac{iQ}4)}
\end{equation}
which is an even function of $x$.
One can show that it obeys a difference equation
\begin{equation}
 \left[\textstyle
  \cosh\pi b(x+\frac m2+\frac{iQ}4)e^{-\pi bp}
 -\cosh\pi b(x-\frac m2-\frac{iQ}4)e^{ \pi bp}\right]
 F_{m,b}(x)~=~0,
\end{equation}
where $p\equiv -\frac i{2\pi}\partial_x$, and a similar equation
with $b$ replaced by $1/b$.
Remarkably, they are invariant if $x,p,m$
are mapped to $p,x,-m$.
This implies the invariance of $F_{m,b}$ under Fourier transform
\cite{BT}
\begin{equation}
 \int dx e^{-2\pi ipx}F_{m,b}(x) ~=~ s_b(m)F_{-m,b}(p).
\end{equation}
It is now easy to show that our partition function for mass-deformed
$T[SU(2)]$ theory is invariant under the exchange of $\mu$ and $\zeta$
if $m$ is sign-flipped at the same time.
\begin{eqnarray}
Z_{\rm 3D}^{{\cal N}=2^\ast}
 &=&
 \frac1{s_b(m)}
 \int d\sigma F_{m,b=1}(\sigma+\mu)F_{m,b=1}(\sigma-\mu)e^{4\pi i\sigma\zeta}
 \nonumber \\ &=&
 \frac1{s_b(-m)}
 \int d\tilde\sigma F_{-m,b=1}(\tilde\sigma+\zeta)
                    F_{-m,b=1}(\tilde\sigma-\zeta)e^{4\pi i\tilde\sigma\mu}.
\end{eqnarray}
Recall that the overall factor $1/s_b(p_e)$ in (\ref{teschner}) played an
important role when we saw that the S-duality kernel reduces
to modular S-matrix element as the external momentum is turned off.
Here this factor is necessary for the 3D partition function
to be precisely invariant under the mirror transformation.
Therefore, this prefactor should arise from the path integral of
some fields except the four charged chiral matters.
Again, we claim it is the determinant of the neutral chiral field $\phi$.

\section{Generalization to $SU(N)$}

Our result can be immediately generalized to the ${\cal N}=2^\ast$
theories with other gauge groups $G$. The theory on the S-duality wall
should be given by $T[G]$ with a real mass deformation. The mass should
be turned on by gauging a $U(1)$ symmetry which commutes with the
global symmetry $G\times G^L$ and flips sign under the mirror
transformation. The anti-diagonal combination of $U(1)_N$ and $U(1)_R$
is the only candidate which meets all these requirements.

It is straightforward to apply the result of \cite{KWY1, KWY2}
to compute the partition function of mass-deformed $T[SU(N)]$ theory
on the S-duality wall $S^3$.
The saddle points of path integral are labelled by
the Coulomb branch parameters $\vec\sigma_{n}=(\sigma_{n,i})$, where
$n=1,\cdots,N-1$ refers to the gauge group $U(n)$ and $i=1,\cdots,n$.
The Coulomb branch parameters of the bulk ${\cal N}=2^\ast$
theories on the two sides of the wall appear as the FI
parameter $\vec\zeta$ and the masses $\vec\mu$ for $N$ fundamental
hypermultiplets.
In the brane picture of Figure \ref{tg}, $\zeta_i$ are identified
with the positions of NS5-branes in the $x_3$-direction and
$\mu_i$ with the position of D5-branes in the $x_7$-direction.
We also require
\[
 \sum_i\zeta_i=\sum_i\mu_i=0.
\]

The partition function reads
\begin{eqnarray}
 Z_{\rm 3D}^{{\cal N}=2^\ast} &=&
 \int \prod_{n=1}^{N-1}d^n\vec\sigma_{n}\
 \prod_{n=1}^{N-1}Z_{V_n}(\vec\sigma_{n})Z_{\phi_n}(\vec\sigma_{n})
 e^{2\pi i\sum_i(\zeta_n-\zeta_{n+1})\sigma_{n,i}}
 \nonumber \\ && \hskip20mm\cdot
 \prod_{n=1}^{N-2}Z_{q_n}(\vec\sigma_n,\vec\sigma_{n+1})
 \cdot Z_{q_{N-1}}(\vec\sigma_{N-1},\vec\mu),
\end{eqnarray}
where $Z_{V_n}$ and $Z_{\phi_n}$ are the one-loop determinants
of the ${\cal N}=4$ $U(n)$ vectormultiplet,
\begin{eqnarray}
 Z_{V_n}(\vec\sigma_n) &=&
 \prod_{i<j}^n\sinh^2\pi(\sigma_{n,i}-\sigma_{n,j}),
 \nonumber \\
 Z_{\phi_n}(\vec\sigma_n) &=&
 \prod_{i,j=1}^ns_{b=1}(\sigma_{n,i}-\sigma_{n,j}-m),
\end{eqnarray}
and $Z_{q_n}$ is that of (bi-)fundamental matters,
\begin{equation}
 Z_{q_n}(\vec\sigma_n,\vec\sigma_{n+1})~=~
 \prod_{i=1}^{n}\prod_{j=1}^{n+1}F_{m,b=1}(\sigma_{n,i}-\sigma_{n+1,j}).
\end{equation}
We expect this partition function to coincide with the S-duality
kernel of Toda conformal blocks on one-punctured torus with
external momentum $m$.

Here we recall that general non-degenerate representations of ${\cal W}_N$
algebra are labeled by the Toda momenta which have $N-1$ components.
Our Toda conformal blocks, on the other hand, have a puncture labelled
by only one mass parameter $m$.
The punctures in generalized quiver gauge theories have been classified
and their relation to Toda vertex operators have been explained in
\cite{Ga, Wy, KMST}.
Each puncture in $SU(N)$ generalized quiver theory is labelled
by a Young diagram with $N$ boxes, which in turn determines
the global symmetry and mass parameters of the theory associated to that
puncture.
In this classification, ${\cal N}=2^\ast$ theories correspond
to a torus with one {\it simple puncture} carrying $U(1)$ global symmetry.

To obtain a fully general formula for the S-duality kernel of
Toda conformal blocks on one-punctured torus, one needs a
generalized quiver gauge theory for a torus with one full
puncture. Such theories are not connected with ${\cal N}=2^\ast$
theories in a simple manner.
No Lagrangian description is available, although one can construct it
from the theory $T_N$, corresponding to the sphere with three full
punctures, by coupling the two $SU(N)$ symmetries to a single
vectormultiplet.

\section{Conclusion}

In this letter we have shown the agreement between the partition function
of mass-deformed $T[SU(2)]$ theory and the S-duality kernel for Liouville
torus one-point conformal blocks. Our result indicates that the AGT
relation can be extended to include the S-duality domain walls.
As suggested in \cite{DGG}, this agreement may be understood in terms
of M5-branes wrapped on $S^3\times {\cal M}_3$, where ${\cal M}_3$
is a three-sphere with a web of defect lines made by connecting
the external legs of two Moore-Seiberg graphs related to each other
by S-duality.

We notice that the evaluation of partition functions
in \cite{KWY1, KWY2} have been performed on a round $S^3$.
It would be a reasonable guess that the one-loop determinants
on a squashed $S^3$ are given by a function $s_b$ with $b\ne1$.
Also, to show the precise matching between the 3D partition function
and the S-duality kernel, we need a full understanding of the
contribution of chiral matters with non-canonical R-charge assignments.
These are interesting future problems.

Very little is known about the three-dimensional theories on the
boundary or domain walls of four-dimensional $N=2$ gauge theories.
It would be an interesting problem to classify such theories and
study their properties under S-duality maps.

{\bf Note added in proof:} Our conjecture about the one-loop determinant for chiral matters
with R-charge 1 was confirmed in the paper by Jafferis \cite{Jafferis} 
and also independently in \cite{Hoso}.

\section*{Acknowledgments}

We thank Seok Kim, Kimyeong Lee, Sangmin Lee, Kazunobu Maruyoshi,
Chaiho Rim, Masato Taki and Piljin Yi for useful discussions.
We also thank the JHEP referee for an important comment on Section
\ref{sec:self-mirror}.
JP appreciates APCTP for its stimulating environement for research.
The work of JP is supported in part by KOSEF Grant R01-2008-000-20370-0,
by the National Research Foundation of Korea(NRF) grant funded by
the Korea government(MEST) with the grant number 2009-0085995
and by the grant number 2005-0049409 through the Center for Quantum
Spacetime(CQUeST) of Sogang University.

\end{document}